\documentstyle[aps,multicol,pra,epsfig]{revtex}

\begin{document}

\draft

\title{Inhomogeneous Superconductivity
in Comb-Shaped Josephson Junction Networks}

\author{P. Sodano$^{1,2}$, A. Trombettoni$^2$}
\address{$^1$ Progetto Lagrange, Fondazione C.R.T. e Fondazione I.S.I.
c/o Dipartimento di Fisica-Politecnico di Torino,
Corso Duca degli Abruzzi 24, Torino, I-10124, Italy \\
$^2$ Dipartimento di Fisica and Sezione I.N.F.N., Universit\`a di
Perugia, Via A. Pascoli, Perugia, I-06123, Italy}

\author
{P.Silvestrini$^{3,4}$, R. Russo$^{3,4}$, and B. Ruggiero$^{4,3}$}
\address{$^3$ Dipartimento d'Ingegneria dell'Informazione,
Seconda Universit\'a di Napoli, Aversa, Italy \\
$^4$ Istituto di Cibernetica ``E. Caianiello'' del CNR, Pozzuoli,
Italy}

\date{\today}
\maketitle

\begin{abstract}
We show that some of the Josephson couplings of junctions arranged
to form an inhomogeneous network undergo a non-perturbative
renormalization provided that the network's connectivity is
pertinently chosen. As a result, the zero-voltage Josephson critical
currents $I_c$ turn out to be enhanced along directions selected by
the network's topology. This renormalization effect is possible only
on graphs whose adjacency matrix admits an hidden spectrum (i.e. a
set of localized states disappearing in the thermodynamic limit). We
provide a theoretical and experimental study of this effect by
comparing the superconducting behavior of a comb-shaped Josephson
junction network and a linear chain made with the same junctions: we
show that the Josephson critical currents of the junctions located
on the comb's backbone are bigger than the ones of the junctions
located on the chain. Our theoretical analysis, based on a discrete
version of the Bogoliubov-de Gennes equation, leads to results which
are in good quantitative agreement with experimental results.
\end{abstract}

\pacs{PACS: 74.81.-g, 74.20.Mn, 71.10.Fd}

%%%%%%COMMENT LINE BELOW FOR 1 COLUMN %%%%%%%%%%%%%%%%%%%%%%%%
%]

\begin{multicols}{2}

It is a common belief that Josephson Junction Networks (JJN) may be
regarded as the prototype of a complex physical system with a
variety of interesting physical behaviors, adjustable acting only on
a few external parameters and, by means of the modern fabrication
technologies, also on the building topology and geometry of the
array \cite{moshe}. It is now possible to build arrays with very
small junctions \cite{naka}
to detect effects due to single electrons in a range
of temperatures related to another relevant energy scale: the
charging energy. The possibility of controlling
experimentally the competition between Josephson and charging
energies makes JJNs useful devices to investigate quantum phase
transitions \cite{sac}, or to model the physical properties of some
real materials like granular superconductors \cite{sima}. Many of
the results valid for JJNs are shared by cold atoms in optical
lattices \cite{anka} since, in these systems, bosonic Josephson
junctions and arrays may be rather easily realized
\cite{catatrombe}; in addition, Josephson networks and devices pave
a very promising avenue to the quantum engineering of states
relevant for quantum computing \cite{MSS}.

Inhomogeneous superconducting networks have been studied
\cite{inhom} mainly to provide a better understanding of the
properties of well controlled disordered granular superconductors
\cite{deut}. The appealing perspective to realize devices for the
manipulation of quantum information recently stimulated the analysis
of inhomogeneous planar JJNs with non conventional connectivity
\cite{ioffe}, engineered to sustain a topologically ordered ground
state \cite{wen}. Transport measurements on superconducting wire
networks evidenced - in a pure system with non-dispersive
eigenstates- interesting anomalies of the network critical current
induced by the interplay between the network's geometry and topology
and an externally applied magnetic field \cite{abilio}; more
recently, the theoretical analysis of rhombi chains has evidenced
the exciting possibility of being able to detect $4e$
superconductivity through measurements of the supercurrent in
presence of a pertinent external magnetic field \cite{feigel}.

In this Letter we show that, even in absence of an externally
applied magnetic field, a JJN fabricated on a pertinent graph
\cite{graf} may support anomalous behaviors of the Josephson
critical currents, which are induced by a non-perturbative
renormalization of some of the Josephson couplings of the array. Our
analysis clearly evidences that this renormalization is only
attainable for the class of graphs, whose adjacency matrix supports
an hidden spectrum \cite{graf,casso}; thus, our findings are not
generic to any inhomogeneous JJN. For instance, in absence of an
external magnetic field, the networks analyzed in
\cite{inhom,deut,ioffe,abilio,feigel} should not give rise to any of
the anomalous behaviors of $I_c$ discussed in this paper.

In the following, we provide a theoretical and experimental study of
the behavior of the Josephson critical currents measurable in a
comb-shaped JJN made of $Nb$ grains located at the vertices of a
"comb" graph and linked by Josephson junctions [see Fig.1]. We
compare our results with those obtained for a linear Josephson
junction chain fabricated with the same junctions. Since one may
regard the backbone of a comb graph as a decorated chain, it appears
natural to compare its superconducting properties with those of a
linear chain since the latter is the simplest network with euclidean
dimension one. The result of this comparison shows that the
Josephson critical currents of the junctions located on the comb's
backbone are sensibly bigger than the ones of the junctions located
on the chain.

Another way to look at a comb-shaped JJN is to regard it as a linear
chain immersed in an environment mimicked by the addition of the
fingers \cite{schmi}. As in many Josephson devices one should then
expect that the nominal value of the Josephson energy $E_J$ of the
junctions in the array gets renormalized by the interaction with the
environment. This situation is often analyzed using either the
Caldeira-Leggett \cite{cale} or the electromagnetic environment
\cite{elen} models. In these approaches one usually assumes that the
{\em effective} boundary conditions for the quantum fluctuations of
the environment modes do not depend on the Josephson couplings or on
the network's topology: while this assumption is perfectly
legitimate for weak environmental fluctuations, better care should
be used if these fluctuations are strong as it may well happen for
one dimensional JJNs. A simple paradigmatic example of a non
perturbative renormalization of Josephson couplings is given by the
simple inhomogeneous one-dimensional array analyzed in
\cite{glala,giuso}, where the source of inhomogeneity is given by
putting on a site of the linear chain a {\em test} junction with a
different nominal value of the Josephson coupling $E_J$. In the
sequel we show that, for a comb-shaped JJN, the Josephson couplings
on the backbone get renormalized. Our explicit computation shows
that this renormalization is indeed non perturbative since the
peculiar connectivity of a comb modifies the spectrum of quantum
modes living on linear chains by the (obviously non-perturbative)
addition of an infinite set of localized states, which disappear in
the thermodynamic limit (the hidden spectrum): adding the fingers to
a backbone chain is, in fact, a topological {\em operation} since it
amounts to a non trivial change of boundary conditions for the
Josephson linear chain. In a different context, the interplay
between an hidden spectrum and a change in boundary conditions has
been recently used in \cite{iofe}.

We use the lattice Bogoliubov-de Gennes (LBdG) equations
\cite{degennes} to compare the properties exhibited by Josephson
linear chains and comb-shaped Josephson networks. Using the
eigenfunctions of the LBdG equations, a self-consistent computation
yields for both systems the gap function, the chemical potential and
the quasi particle spectrum. We show that, for a linear chain, the
superconducting gap and critical temperature satisfy to the
well-known BCS equations and that, on the backbone of a comb JJN,
the BCS equations are satisfied with a renormalized value of the
Josephson energy. Then, we compute the zero-voltage Josephson
critical currents $I_c$ on the comb's backbone and compare our
results for $I_c$ with the outcomes of experimental measurements:
our computation not only confirms with good accuracy the
experimental results of \cite{exper}, but is also in good agreement
with new data obtained at temperatures closer to the critical
temperature for the onset of superconductivity in $Nb$ grains. The
new data are shown in Fig.2.

To obtain a discrete version of the BdG equations suitable to
describe the JJNs fabricated in \cite{exper}, we make the ansatz
that the eigenfunctions of the continuous BdG equations
\cite{degennes} may be written in a tight binding form as
$u_\alpha(\vec{r}) = \sum_i u_\alpha(i) \phi_i(\vec{r})$ and
$v_\alpha(\vec{r}) = \sum_i v_\alpha(i) \phi_i(\vec{r})$; $i$ labels
the position of a superconducting island while the contribution of
the electronic states participating to superconductivity in a given
island is effectively described by a field $\phi_i(\vec{r})$, whose
specific form depends on the geometry of the islands and on the
fabrication parameters of the connecting junctions. The LBdG
equations then read
\begin{equation}
\in_\alpha u_\alpha(i) = \sum_j \epsilon_{ij} u_\alpha(j) +
\Delta(i) v_\alpha(i) \label{LBdG1}
\end{equation}
\begin{equation}
\in_\alpha v_\alpha(i) = - \sum_j \epsilon_{ij} v_\alpha(j) +
\Delta^{\ast}(i) u_\alpha(i). \label{LBdG2}
\end{equation}
where $u_\alpha$ and $v_\alpha$ satisfy to $\sum_i \left[ \vert
u_{\alpha}(i)\vert^2 + \vert v_{\alpha}(i)\vert^2 \right]=1$. The
matrix $\epsilon_{ij}$ is defined by $ \epsilon_{ij}=-t A_{ij} +
U(i) \delta_{ij} - \tilde{\mu} \delta_{ij}$, with $A_{ij}$ being the
adjacency matrix of the network \cite{noi1}, $\tilde{\mu}=\mu-\int
d\vec{r} \phi_i(\vec{r}) \left( -\hbar^2 \nabla^2 / 2m \right)
\phi_i(\vec{r})$ and $t = - \int d\vec{r} \phi_i(\vec{r}) [ -\hbar^2
\nabla^2/2m + U_0(\vec{r})] \phi_j(\vec{r}) \approx E_J$.
$E_J=(\hbar/2e)I_c$ is the nominal value of the Josephson energy of
all the junctions in the network while $I_c$ is the unrenormalized
zero-voltage Josephson critical current of each junction.
$U_0(\vec{r})$ mimics the effects of the barrier between the
superconducting islands. Self-consistency requires $
\Delta(i)=\tilde{\cal V} \sum_\alpha  u_\alpha(i)
v^{\ast}_{\alpha}(i) \tanh{\left( \frac{\beta}{2} \in_\alpha
\right)}$ and $U(i)=-\tilde{\cal V} \sum_\alpha \left[ \vert
u_\alpha(i) \vert^2 f_\alpha + \vert v_\alpha(i) \vert ^2 \left(
1-f_\alpha \right) \right]$, where $\tilde{\cal V} \equiv {\cal V}
\phi^2(\vec{r}=\vec{r}_i)$ is assumed to be independent on $i$.
Topology is encoded in the term $-t A_{ij}$ appearing in the
definition of the matrix $\epsilon_{ij}$, while the specific values
of $t$ and $\tilde{\cal V}$ depend - as a result of our ansatz on
the form of the eigenfunctions of the BdG equations- only on the
$\phi_i(\vec{r})$.

\begin{figure}
\centerline{\psfig{figure=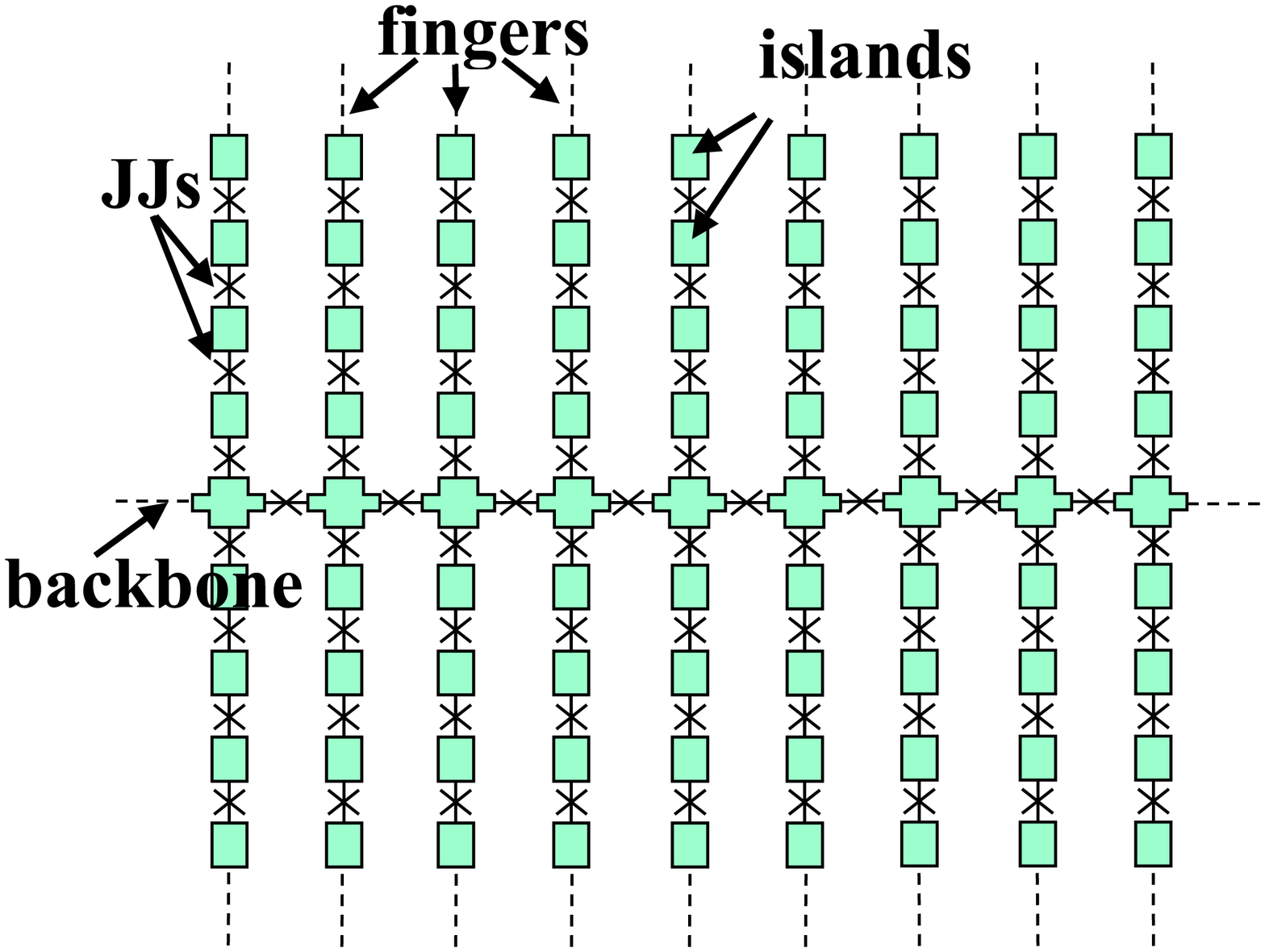, width=60mm,angle=270}}
\caption{Schematic drawing of a comb array. The superconducting
islands (full box) are connected in series to each other through
Josephson junctions (JJs). The finger arrays are connected to each
other only through JJs to the central islands forming the backbone
array.} \label{comb}
\end{figure}

To justify the assumptions involved in the derivation of
Eqs.(\ref{LBdG1})- (\ref{LBdG2}), we observe that, for the JJNs
described in \cite{exper}, capacitive (inter islands and with a
ground) effects are negligible, that the total number of electrons
on the island ${\cal N}$ is much larger than the number of electrons
tunneling through the Josephson junction and that all islands
contain approximately the same ${\cal N}$ (${\cal N}(i) \equiv {\cal
N}$). Furthermore, the islands are big enough to support the same
superconducting gap of the $Nb$ bulk material. As a result one may
require $\phi_i(\vec{r})$ to be position-independent on each island
except for a small region near the junction and to be the same on
each island with a normalization given by $\int d\vec{r}
\phi_i(\vec{r}) \phi_i(\vec{r}) = {\cal N}(i) \equiv {\cal N}$ and
$\int d\vec{r} \phi_i(\vec{r}) \phi_j(\vec{r}) \approx 0$ for $ i
\neq j$. In our derivations we put ${\cal N} \equiv 1$.

For a linear array, the LBdG may be readily solved leading to an
uniform potential $U(i) \equiv U_c$ and an uniform pair potential
$\Delta(i) \equiv \Delta_c$. From the eigenvalue equation $-E_J
\sum_j A_{ij} \psi_k(j) = e_k \psi_k(i)$, one gets $e_k=-2 E_J
\cos{k}$: it follows $\in_k=\sqrt{\Delta_c^2+E_k^2}$ with
$E_k=e_k+U_c-\tilde{\mu}$. The BCS-like behaviour is obtained when
$E_k = e_k$, which happens since $U_c = 0$ and ${\mu}=E_F$. When
$\Delta_c / E_J \ll 1$, for $T=0$, one gets $ \Delta_c(T=0) = 8 E_J
e^{-2 \pi E_J / \tilde{\cal V}}$, while, for $T=T_c$ (i.e.,
$\Delta_c(T=T_c)=0$), one obtains $k_B T_c={\cal C} E_J e^{-2 \pi
E_J / \tilde{\cal V}}$, with ${\cal C} = 4.54$. It is comforting
that the assumptions on which our approach is based lead, for the
chain, to results having the same functional form of the well-known
BCS formulas for the gap at $T=0$ (i.e., $\Delta(T=0)=2 \hbar
\omega_D e^{-1/n(0) V_{BCS}}$) and the BCS critical temperature
(i.e., $k_B T_c=1.14 \hbar \omega_D e^{-1/n(0) V_{BCS}}$), provided
that $n(0) V_{BCS} \ll 1$ \cite{degennes}: in addition, one gets
also $\Delta_c(T=0) / k_B T_c=8 / {\cal C} \approx 1.76$.

Measurements on a chain made with $Nb$ grains yield $T_c \approx 8.8
K$ and $\Delta_c(T=0) \approx 1.4 meV \approx k_B \cdot 15.9 K$;
furthermore, in the experimental setup described in \cite{exper} it
is $I_c \approx 18 \mu A$. The parameters $E_J$ and $\tilde{{\cal
V}}$, determined from the BCS equation yielding the chain's critical
temperature, are then given by $E_J \approx k_B \cdot 430 K$ and
$\tilde{{\cal V}}/ E_J =1.185$. In Fig.2 we plot for several
temperatures the measured $I_c$ (circles) and the critical currents
obtained inserting $\Delta_c(T)$ in the well-known
Ambegaokar-Baratoff expression \cite{amba} for the zero-voltage
Josephson current (lower solid curve): the agreement is excellent.

For a comb network with $N \times N$ islands (see Fig.1), one finds
a solution of the LBdG equations (\ref{LBdG1})- (\ref{LBdG2}) where
both the Hartree-Fock potential $U(i)$ and the gap function
$\Delta(i)$ are position dependent. We denote the islands by
$(x,y)$, $x$ labeling the finger and $\vert y \vert$ the distance
from the backbone, expressed in lattice units. The eigenvalue
equation $-E_J\sum_j A_{ij} \psi_\alpha(j) = e_\alpha
\psi_\alpha(i)$, admits \cite{noi1}, in addition to a set of
delocalized states with energies ranging from $-2E_J$ to $2E_J$, a
localized ground-state $\psi_0 = (C_0 / \sqrt{N}) e^{- y /\xi}$,
corresponding to the eigenvalue $e_0=-2 \sqrt{2} E_J$
($C_0^2=1/\sqrt{2}$ and $\xi$ given by $\sinh{(1/\xi)}=1$) and an
hidden spectrum made of other eigenstates localized around the
backbone \cite{noi1}. For a crude analytical estimate, one may
require that, away from the backbone, the fingers may be regarded as
a linear chain with uniform potentials (i.e., $\Delta(i)=\Delta_c$
and $U(i)=U_c$). To get then coupled equations for $\Delta_b$,
$\Delta_c$, $U_b$, and $U_c$, one writes the LBdG equations
(\ref{LBdG1})-(\ref{LBdG2}) on a backbone's grain $i$. We set
$u_\alpha(i)=U_\alpha \psi_\alpha(i)$ and $v_\alpha(i)=V_\alpha
\psi_\alpha(i)$, with $U_\alpha^2+ V_\alpha^2=1$. The
self-consistency equation for $U$ implies that, at $T=0$, $U_b
\approx U_c - \frac{\tilde{\cal V} C_0^2}{2}$; upon requiring
$\tilde{\mu} \approx U_b $ one immediately sees that, due to the
localized modes in the fermionic spectrum, the chemical potential on
the comb's backbone is smaller than the one measured on the chain.

By substituting the wavefunctions of the eigenstates of the hidden
spectrum \cite{noi1} in Eqs.(\ref{LBdG1})-(\ref{LBdG2}) and
using $ \tilde{\mu} \approx U_b $ one gets
\begin{equation}
\Delta_b=\Delta_c+ \frac{\Delta_b \tilde{\cal V}}{\pi} \cdot
\int_{0}^{\pi/2} dk \frac{\cos{k}}{\in_k \sqrt{1+\cos^2{k}}} \cdot
\tanh{ \left( \frac{\beta}{2} \in_k \right) }. \label{D-U-1}
\end{equation}
where $\in_k=\sqrt{\Delta_b^2 + 4 E_J^2 \left( 1+\cos^2{k}
\right)}$. The hidden spectrum eigenstates contribute to the gap
function $\Delta_b$ through the second term in the rhs of
Eq.(\ref{D-U-1}): without them, $\Delta_b$ equals $\Delta_c$.

When $E_J \gg \Delta_b, \Delta_c$, Eq.(\ref{D-U-1}), at $T=0$,
yields $ \frac{\Delta_b(T=0)}{\Delta_c(T=0)}=
\frac{1}{1-\frac{\eta_c \tilde{\cal V}}{2 \pi E_J} } \equiv {\cal
K}$ where $\eta_c \equiv (1/\sqrt{2}) \, \log{ \left( 1+\sqrt{2}
\right) }$. Furthermore, at low temperatures, $\Delta_b(T) /
\Delta_c(T) \approx \Delta_b(T=0) / \Delta_c(T=0)$. Using the
parameters $E_J$ and $\tilde{\cal V}$ obtained from the measurements
carried on the JJ chain, for a JJ comb one gets ${\cal K} \approx
1.13$.

Upon requiring that, as for the linear chain, the $T=0$ backbone's
gap function has a BCS like functional form, i.e. $\Delta_b(T=0)=8
\bar{E}_J e^{ -2 \pi \bar{E}_J / \bar{ {\tilde{\cal V}} } }$, with
$\bar{E}_J$ and $\bar{ {\tilde{\cal V}} }$ the renormalized
Josephson energy and the renormalized interaction term, one is able
to estimate the renormalization of the Josephson coupling within the
LBdG approach. Namely, one has,
\begin{equation}
\bar{E}_J={\cal K} E_J; \, \, \, \, \, \, \, \, \, \bar{
{\tilde{\cal V}} }={\cal K} {\tilde{\cal V}}, \label{ren}
\end{equation}
which embodies the effects of the hidden spectrum on the Josephson
couplings.

In Fig.2  we plot, as a function of the normalized temperature, the
values of $I_c$ measured with the methods described in \cite{exper}
(squares) and the values of $I_c$ obtained from the
Ambegaokar-Baratoff formula using both the renormalized coupling
given by Eq.(\ref{ren}) and the gap function along the backbone for
the comb-like JJN studied in \cite{exper} (solid curve): the
agreement between theory and experiments is very good at low
temperatures, while the theory gives a slight overestimate at higher
temperature.

\begin{figure}
\centerline{\psfig{figure=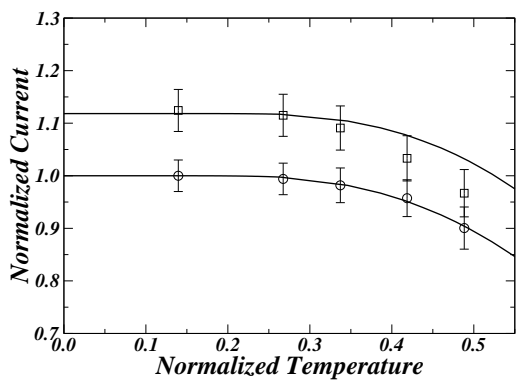, width=60mm, angle=270}}
\caption{Critical currents (in units of the critical current on the
reference chain at $T=1.2 K$) as a function of $T/T_c$ for the
backbone and the chain. The solid lines are the estimated critical
currents for the backbone (top) and the chain (bottom). Circles
(squares): experimental values for the chain (backbone).}
\label{ambbar}
\end{figure}

We showed that a non perturbative (i.e. induced by the states of the
hidden spectrum) renormalization of some of the Josephson couplings
of a comb-shaped JJN is responsible for the observed enhancement of
$I_c$ of the Josephson junctions located along the comb's backbone.
The key assumption in our derivation is that the eigenfunctions of
the BdG equations may be written in a tight binding form; once this
assumption is made, one is able to derive Eqs. (\ref{LBdG1})-
(\ref{LBdG2}) and to account for all the dependence on the
electronic states into the definition of the parameters $E_J$ and
${\tilde{\cal V}}$, which, in this paper, we determined from the
measurements carried on the linear chain. Our approach yields a
value of the renormalized Josephson coupling of the junctions
located on the comb's backbone in excellent agreement with the
experimental results (see fig.2). We expect that similar phenomena
happen for the class \cite{casso} of JJNs fabricated on graphs whose
adjacency matrix supports an hidden spectrum.

We are much grateful to L.B. Ioffe and M. Rasetti for asking very
useful questions at several stages  of our work. We are indebted to
C. Granata, V. Corato and F.P. Mancini, for their helpful
assistance. We thank M. Cirillo for discussions. Our research has
been partially supported by the MIUR Project {\it Josephson Networks
for Quantum Coherence and Information} (grant No.2004027555).

\end{multicols}

\end{document}